\documentclass[lettersize,journal]{IEEEtran}
\usepackage{amsmath,amsfonts}
\usepackage{algorithm}
\usepackage{array}
\usepackage{multirow}
\usepackage{subfigure}
\usepackage{textcomp}
\usepackage{stfloats}
\usepackage{url}
\usepackage{verbatim}
\usepackage{graphicx}
\usepackage{cite}
\usepackage{caption}
\usepackage{amssymb}
\usepackage{algpseudocode}
\usepackage{soul}    
\usepackage{xcolor}  
\usepackage{array}
\usepackage{booktabs} 
\usepackage{colortbl} 
\usepackage{array} 
\definecolor{lightgray}{gray}{0.9} 
\usepackage{booktabs}
\usepackage{siunitx}
\usepackage{multirow}


\begin{document}

\title{-8 dB SNR + 90\% Packet Loss: MamVSC — CSI-Guided Semantic Mamba for Extreme-Robust Video Semantic Communication}
\author{\relax Lei Teng, Senran Fan, Chen Dong\IEEEauthorrefmark{1}, Haotai Liang, Xiaodong Xu, \textit{Senior Member, IEEE}, Ping Zhang, \textit{Fellow, IEEE}
	\thanks{This work was supported in part by the National Key R\&D Program of China No. 2020YFB1806905 and the Beijing Natural Science Foundation No. L251035.\\
		Lei Teng, Senran Fan, Chen Dong, Haotai Liang, Xiaodong Xu and Ping Zhang are with the State Key Laboratory of Networking and Switching Technology, Beijing University of Posts and Telecommunications, Beijing 100876, China. (e-mail: tenglei@bupt.edu.cn; fansenran@bupt.edu.cn;  dongchen@bupt.edu.cn; 	lianghaotai@bupt.edu.cn; xuxiaodong@bupt.edu.cn; pzhang@bupt.edu.cn.)}
}

\markboth{Journal of \LaTeX\ Class Files,~Vol.~14, No.~8, August~2021}%
{Shell \MakeLowercase{\textit{et al.}}: A Sample Article Using IEEEtran.cls for IEEE Journals}

\maketitle

\begin{abstract}

Semantic communication, leveraging joint source-channel coding, is designed to mitigate semantic distortion introduced by the channel. However, most current studies focus solely on semantic deviation distortion caused by physical wireless channels, while overlooking semantic erasure distortion due to packet loss. A CSI-Guided Mamba-based video semantic wireless digital communication system (MamVSC) employing semantic grouping is proposed to simultaneously address both semantic deviation and erasure distortions. In this system, a semantic Mamba module, guided by channel state information (CSI) feedback, is utilized to dynamically adjust the granularity of extracted semantic information, adapting to channel conditions. Furthermore, a Semantic Channel Codec based on dynamic Semantic clustering centers is introduced, where the distance between semantic vectors within the same semantic class and their corresponding Semantic clustering center is dynamically adjusted according to channel conditions, enhancing robustness against channel noise. Additionally, a adaptive packet loss recovery module, dynamically adaptive to the CSI, is proposed. The system achieves an MS-SSIM greater than 0.6 and a PSNR exceeding 21 dB at an SNR of -8 dB and a packet loss rate of 90\% in AWGN channel.

\end{abstract}


\section{Introduction}

\IEEEPARstart{A}{s} the physical-layer performance of traditional communication systems approaches the Shannon limit while user demands for communication continue to escalate, emerging technologies are imperative to further improve transmission efficiency under constrained time–frequency resources \cite{ref1}. Moreover, the conventional lossless transmission paradigm exhibits a pronounced cliff effect in harsh communication environments, severely degrading system reliability. These limitations indicate that further performance gains can hardly be achieved through physical-layer optimization alone, thereby calling for a paradigm shift that rethinks communication from an information-processing perspective.

Driven by this need, recent years have witnessed a deep convergence of communication and artificial intelligence (AI). Within the IMT-2030 framework, AI and Communication has been identified as a key enabling direction for enhancing overall communication capabilities through AI-driven technologies \cite{ref2}. Despite their diverse forms, these AI-empowered paradigms share a common objective: transcending conventional bit-level transmission to achieve more efficient and intelligent information delivery. In this context, semantic communication has emerged as a particularly promising paradigm.

One year after Shannon introduced information theory \cite{ref3}, Weaver conceptualized communication into three hierarchical levels: syntax, semantics, and pragmatics \cite{ref4}. Traditional communication systems have long been confined to the syntactic level, emphasizing bit-accurate transmission. However, propelled by rapid advances in artificial intelligence, semantic-level communication has recently attracted significant research attention \cite{ref5,ref6}. By extracting and transmitting task-relevant semantic information from the intrinsic meaning of the source, semantic communication substantially reduces communication overhead while improving transmission efficiency. Furthermore, semantic communication inherently exhibits greater tolerance to bit-level errors, thereby offering superior robustness under adverse channel conditions \cite{ref7,Liang2025a}. These advantages make semantic communication particularly attractive for data-intensive and perception-oriented applications.

Among such applications, video stands out as one of the most data-intensive and semantically rich media modalities, rendering it both a promising and challenging scenario for semantic communication. In practical wireless video transmission, two primary forms of semantic distortion commonly arise: \emph{semantic deviation}, mainly induced by channel noise, fading, and interference, and \emph{semantic erasure}, predominantly caused by packet losses in packet-switched systems. While extensive efforts have been devoted to video semantic communication, most existing schemes are designed to address only one of these distortions in isolation. As a result, achieving comprehensive robustness against both semantic deviation and semantic erasure in practical wireless video transmission remains an open challenge.

To address this challenge, this paper proposes a Mamba-based video semantic communication system that jointly mitigates semantic deviation and semantic erasure. By adopting Vision Mamba (VIM) \cite{ref30,Gu2023} as the backbone—owing to its strong global modeling capability—we further introduce semantic grouping guided by dynamic semantic clustering centers, along with CSI-adaptive packet loss recovery mechanisms. The proposed system enables robust video reconstruction even under severe channel impairments.

The main contributions of this paper are summarized as follows:
\begin{enumerate}
	\item A Mamba-based video semantic digital communication system is proposed, which employs semantic grouping and channel state information (CSI) feedback to simultaneously mitigate semantic deviation and semantic erasure. A Semantic Mamba Block based on the CSI-Guided Attentive State-Space Equation is designed.
	\item A semantic channel codec based on dynamic semantic clustering centers is introduced to enhance intrinsic robustness.
	\item A packet loss recovery module that dynamically adapts to CSI is proposed for effective semantic erasure compensation.
	\item Extensive simulations and ablation studies validate the superior performance and effectiveness of each proposed module.
\end{enumerate}
The remainder of this paper is organized as follows: Section II introduces the related work, Section III presents the system model, Section IV details the proposed MamVSC framework, Section V provides simulation results and comparisons, and Section VI concludes the paper.

\section{System Model}
\begin{figure}[!htbp]
	\centering
	\includegraphics[width=0.5\textwidth]{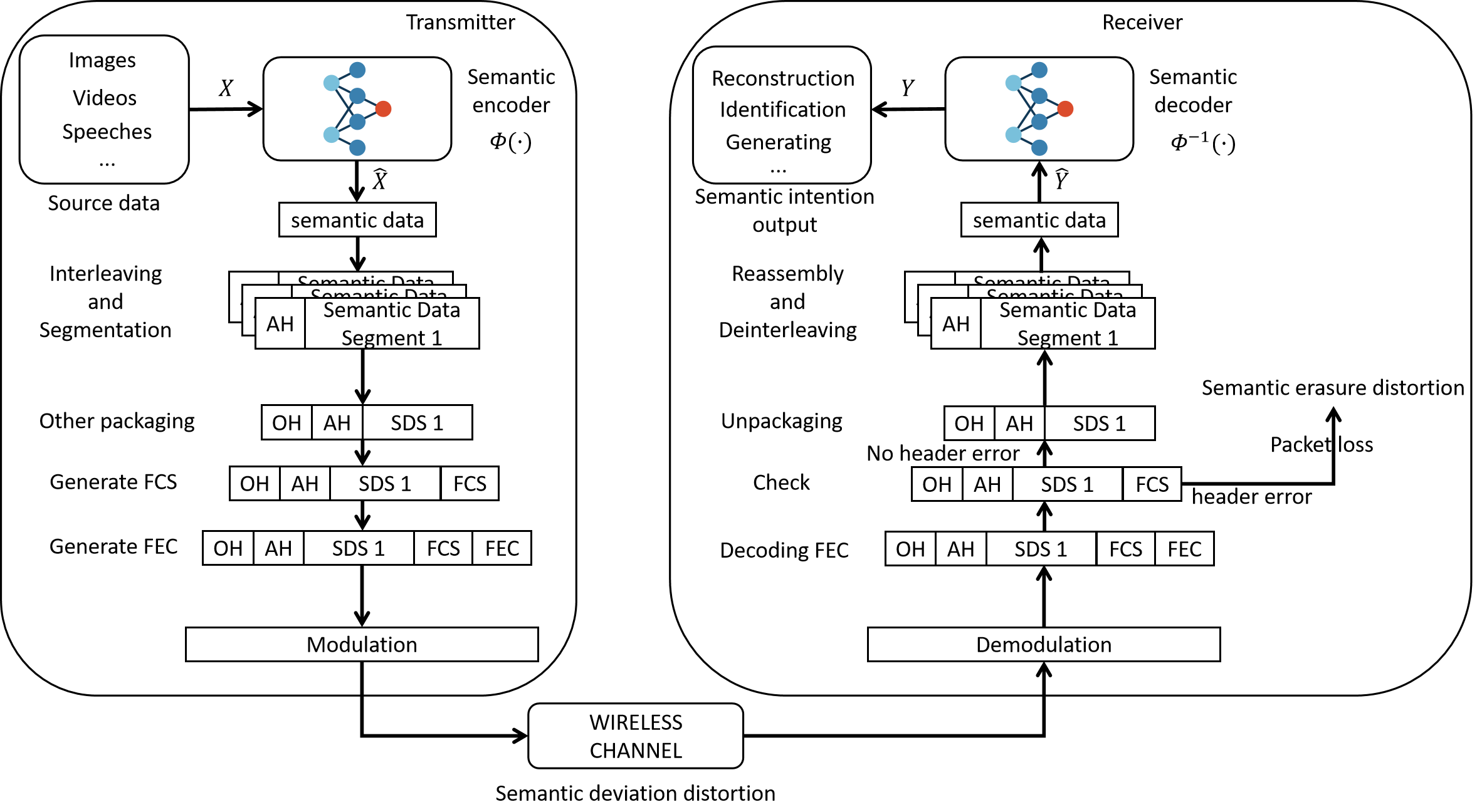}\\
	\caption{\footnotesize The architecture diagram of the semantic digital communication system is presented, illustrating the two sources of semantic distortion in wireless communication: the wireless channel and packet loss. At the transmitter, source data is processed by a semantic encoder to extract semantic information. This information undergoes semantic-level interleaving and segmentation, with an application-layer header (AH) added as described in \cite{ref27}. Subsequently, other encapsulation steps append other headers (OH). Finally, a frame check sequence (FCS) and forward error correction (FEC) code are added for the header data, and the resulting data is modulated into a digital signal for transmission. During transmission over the wireless channel, semantic deviation occurs. At the receiver, the received signal is first demodulated, followed by channel decoding and header verification. If errors are detected in the header, packet loss occurs, leading to semantic erasure. If no errors are present, the process continues with depacketization, reassembly, and deinterleaving, followed by semantic decoding.}
	\label{system model}
\end{figure}
 
The framework of the proposed semantic digital communication system is illustrated in Fig. 1. At the transmitter, the source data $ X $ is first encoded into semantic data $ \hat{X} $ by a semantic encoder. Subsequently, based on the semantic-level interleaving and segmentation method proposed in \cite{ref27}, the semantic data is divided into multiple semantic data segments and corresponding application layer headers (AH) are added. The AH consists of three bytes: the first byte indicates the current GOP index, while the second and third bytes represent the segment label, denoting the position of the segment within the GOP semantic data. The GOP index enables the receiver to determine whether the received data belongs to the same GOP. Assuming no out-of-order reception, a change in the received GOP index compared to that in the buffer indicates that the semantic data of the previous GOP has been fully received and can be forwarded to the semantic decoder for video decoding. Using segment labels, the segments are reassembled into complete semantic data, with missing segments replaced by zeros, even in the presence of packet loss during transmission. Each segment then undergoes a packetization process, where a Other Header (OH) is added. Following this, a frame check sequence (FCS) and forward error correction (FEC) codes are generated for the header portion. The specific implementation and distinctions between FCS and FEC are elaborated in subsequent subsections. Finally, the packets are digitally modulated and transmitted over a wireless channel to the receiver. During transmission, the semantic data may be affected by channel noise, resulting in semantic deviation distortion.
At the receiver, the received packets are first demodulated to recover the bitstream. The bitstream is then processed through channel decoding, and the FCS is used to verify the integrity of the header information. If the header information is correct, the semantic data segments are extracted through a depacketization process. Subsequently, multiple segments of the same semantic data are reassembled and deinterleaved to restore the original semantic data format. The semantic data is then fed into a semantic decoder to produce the final semantic intent output. If errors are detected in the header during the verification step, the affected semantic data segment cannot proceed through the depacketization process, leading to packet loss and, consequently, semantic erasure distortion.

\begin{figure}[!htbp]
	\centering
	\includegraphics[width=0.49\textwidth]{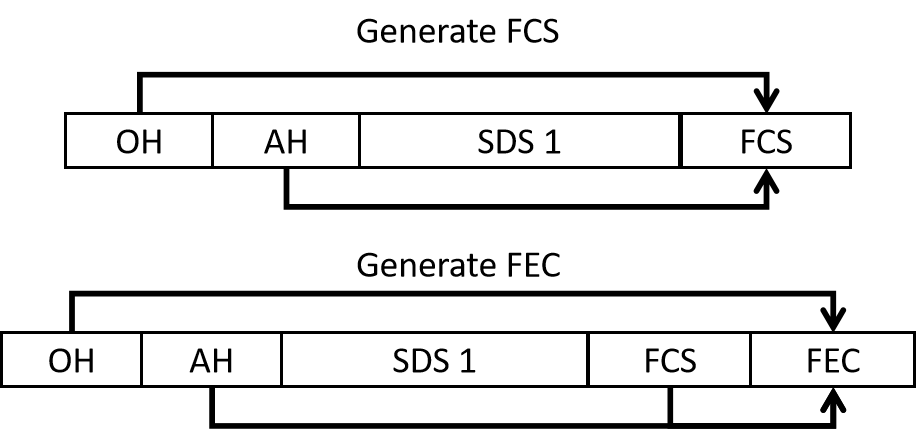}\\
	\caption{\footnotesize Frame Structure Diagram.}
	\label{Frame}
\end{figure}

\textit{Frame Check Sequence and Forward Error Correction}: Due to the characteristics of joint source-channel coding training, semantic communication exhibits high robustness to bit-level errors. Consequently, in the verification process, the stringent requirements for bit errors in semantic data can be relaxed, and independent channel coding is not required. However, packets include not only semantic information but also header information encapsulated according to the protocol, which is highly sensitive to any bit errors. Errors in the header may lead to incorrect destination addresses, potentially causing packets to be misrouted to the wrong receiver, resulting in data leakage or communication disruption. Additionally, errors in protocol fields may trigger exceptions or crashes in the protocol stack. Therefore, in the proposed semantic digital communication system framework, FCS and FEC mechanisms are retained but applied exclusively to the header portion. Specifically, FCS is generated solely based on the header data, and FEC encoding is performed only for the header and FCS portions, thereby maximizing the robustness of semantic communication to bit-level errors.

Furthermore, it is essential to distinguish between the two primary types of semantic distortion introduced by wireless communication: \textit{semantic deviation} and \textit{semantic erasure}.
﻿

\textit{Semantic deviation} arises from channel noise directly corrupting the semantic data payload (SDS in Fig. 2). Since semantic communication adopts joint source-channel coding without separate channel coding for the payload, the transmitted semantic vectors are exposed to noise during propagation over the wireless channel. This results in a continuous perturbation of the received semantic vectors, where the deviation is pervasive across all received information. Such distortion manifests as gradual quality degradation (e.g., noise, blurring, or color shift) in the reconstructed video, but preserves the overall semantic structure, allowing the receiver to still infer the core meaning with partial robustness inherent to semantic representations.
﻿

In contrast, \textit{semantic erasure} is a binary, catastrophic distortion triggered by packet loss. As illustrated in Fig. 1, only the header (OH + AH) is protected by traditional error detection and correction mechanisms. When the number of bit errors in the protected header exceeds the correction capability of the FEC, header verification fails, causing the entire packet to be discarded at the receiver. Consequently, the corresponding semantic data segment is completely lost, introducing ``erasures'' (zero-filled) in the reassembled semantic feature map. This type of distortion leads to abrupt, localized loss of semantic information—such as missing objects, discontinued motions, or large blank regions in certain frames—significantly more destructive than deviation when packet loss rate is high.

The coexistence of both distortions in practical wireless video transmission (especially in low-SNR scenarios) poses a severe challenge: semantic deviation degrades global fidelity, while semantic erasure destroys local completeness. Traditional communication systems address bit errors uniformly, exhibiting a ``cliff effect'' under poor channels. In contrast, semantic communication can tolerate moderate deviation due to its meaning-centric nature, but remains vulnerable to erasure without dedicated recovery mechanisms. Therefore, an effective video semantic communication system must jointly combat both distortions through channel-adaptive semantic encoding and erasure-aware recovery, which is the core objective of the proposed MamVSC framework.

\section{The Proposed MamVSC Framework}

\begin{figure}[!htbp]
	\centering
	\includegraphics[width=0.5\textwidth]{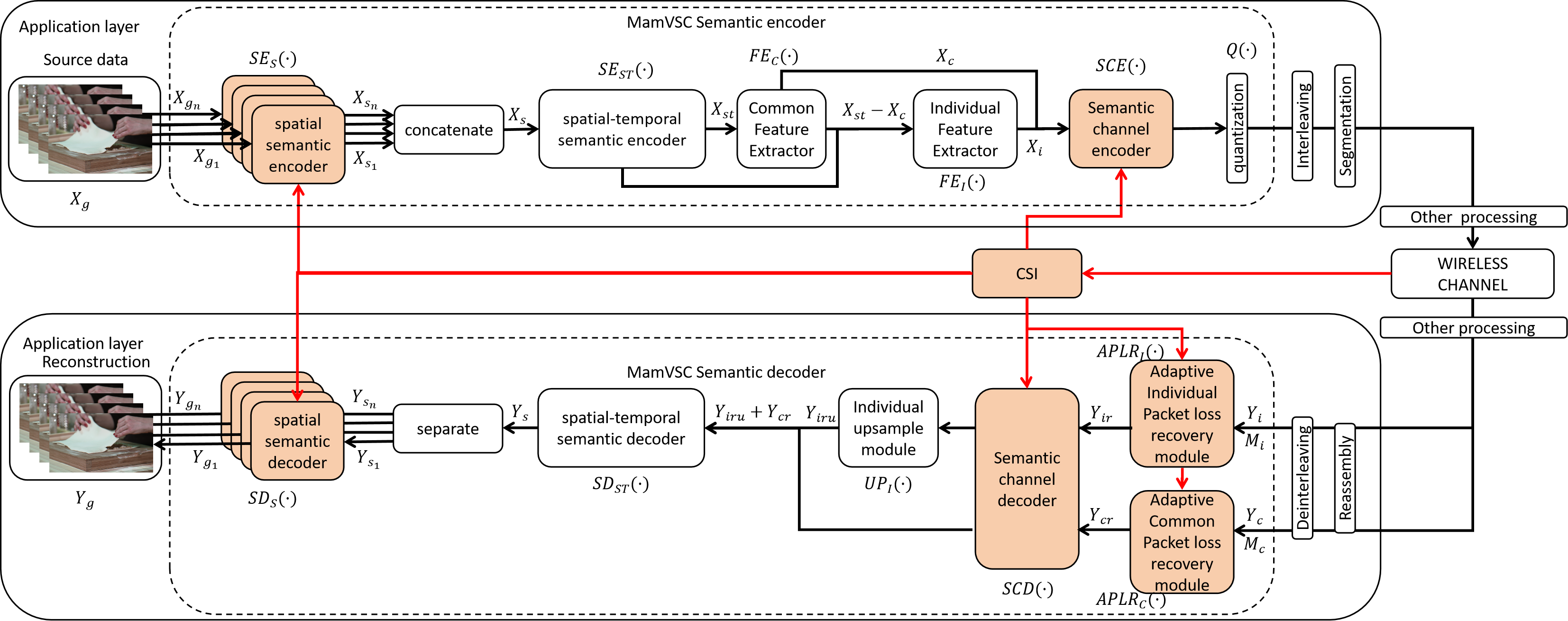}\\
	\caption{\footnotesize The overall architecture of our MamVSC for wireless video transmission. The orange box represents the newly proposed framework in this paper, comprising the Semantic Mamba Block, Semantic Channel Codec, and Adaptive Packet Loss Recovery Module, while the white box denotes the basic modules proposed in our previous article \cite{ref27}.}
	\label{MamVSC}
\end{figure}

The overall framework of the proposed MamVSC system is depicted in Fig. \ref{MamVSC}. At the transmitter, input video data is processed by a semantic encoder, followed by semantic-level interleaving and segmentation\cite{ref27}. Subsequently, the data undergoes additional packetization steps, is modulated into signals, and transmitted over a wireless channel to the receiver. At the receiver, the received signals are demodulated, depacketized, reassembled, and deinterleaved. The semantic codec then processes the semantic information, which may contain semantic deviation and erasure, to produce the reconstructed video. Building on the MSTVSC framework proposed in \cite{ref27}, CSI is incorporated into the codec to enhance robustness against semantic distortions, thereby endowing the model with channel-adaptive capabilities. Additionally, the backbone of the spatial encoder is replaced with the Mamba model to improve reconstruction performance.

Specifically, the encoding and decoding process of MamVSC is described as follows:

Assume the input is a video sequence \( X \) with temporal length \( T \), height \( H \), and width \( W \), represented as:
 \begin{equation}
	\begin{split}
X = \{X_1, X_2, \dots, X_T\},
	\end{split}
\end{equation}

where \( X_t \in \mathbb{R}^{H \times W \times 3} \) denotes the pixel intensity vector of the frame at time step \( t \).

At the transmitter, the video sequence \( X \) is divided into multiple Groups of Pictures (GOPs) \( X_g \), each containing \( n \) frames, serving as a single input for MamVSC. A specific GOP is expressed as:
 \begin{equation}
	\begin{split}
X_g = \{X_{g_1}, X_{g_2}, \dots, X_{g_n}\}, \quad X_{g_j} \in X, \quad j=1,2,\dots,n.
	\end{split}
\end{equation}
Subsequently, each frame \( X_{g_j} (j=1,2,\dots,n) \) in \( X_g \) is sequentially input into the Spatial Semantic Encoder. Unlike \cite{ref27}, CSI is introduced to dynamically adjust the precision of semantic extraction, generating the corresponding spatial semantic information vector 
\( X_{s_j} \):
 \begin{equation}
	\begin{split}
X_{s_j} = \text{SE}_S(X_{g_j}), \quad X_{g_j} \in X_g, \quad j=1,2,\dots,n,
	\end{split}
\end{equation}

where \( \text{SE}_S(\cdot) \) denotes the function of the spatial semantic encoder. The spatial semantic information vectors \( X_{s_j} \) are concatenated to form the spatial semantic information set \( X_s \):
 \begin{equation}
	\begin{split}
		X_s = \{X_{s_1}, X_{s_2}, \dots, X_{s_n}\}.
	\end{split}
\end{equation}
Next, \( X_s \) is input into the Spatial-Temporal Semantic Encoder for temporal semantic information extraction and further spatial semantic compression, producing the spatio-temporal semantic information vector \( X_{st} \):
 \begin{equation}
	\begin{split}
		X_{st} = \text{SE}_{ST}(X_s),
	\end{split}
\end{equation}
where \( \text{SE}_{ST}(\cdot) \) represents the function of the spatio-temporal semantic encoder.

To further compress temporal redundancy within the GOP, \( X_{st} \) is input into the Common Feature Extractor to extract the common temporal feature \( X_c \). The individual feature is obtained by subtracting \( X_c \) from \( X_{st} \), i.e., \( X_{st} - X_c \). This individual feature is then processed by the Individual Feature Extractor  to further extract and compress semantic information, yielding the final individual information \( X_i \):
 \begin{equation}
	\begin{split}
		X_c = \text{FE}_C(X_{st}),
	\end{split}
\end{equation}
 \begin{equation}
	\begin{split}
		X_i = \text{FE}_I(X_{st} - X_c),
	\end{split}
\end{equation}
where \( \text{FE}_C(\cdot) \) and \( \text{FE}_I(\cdot) \) denote the functions of the common and individual feature extractors, respectively. Common features $X_c$ primarily capture slowly varying information across the GOP. In contrast, individual features $X_i$ effectively retain rapidly changing information.

To adapt to varying channel conditions, \( X_c \) and \( X_i \) are input into the Semantic Channel Encoder (SEC), which performs dynamic Semantic clustering center scaling adjustments based on CSI to enhance robustness:
 \begin{equation}
	\begin{split}
		(X_c', X_i') = \text{SEC}(X_c, X_i).
	\end{split}
\end{equation}

Subsequently, a quantization module \( Q(\cdot) \) normalizes the data, compresses the feature channels to a specified number \( C \), and applies rounding quantization to meet digital communication requirements. Finally, following the interleaving and segmentation process described in [27], the data undergoes additional packetization steps and is transmitted over the wireless channel to the receiver.

At the receiver, the received semantic information is first depacketized, followed by reassembly and deinterleaving at the application layer. Using segmentation tags in the application-layer headers, missing semantic information due to packet loss is identified, with missing positions marked as 1 and non-missing positions as 0, generating semantic erasure matrices \( M_c \) and \( M_i \) for the common semantic feature \( Y_c \) and individual semantic feature \( Y_i \), respectively. The pairs \( (Y_c, M_c) \) and \( (Y_i, M_i) \) are input into the Adaptive Common Packet Loss Recovery Module  and Adaptive Individual Packet Loss Recovery Module , respectively, to generate complete common semantic features \( Y_{cr} \) and individual semantic features \( Y_{ir} \) based on un-lost semantic information:
\begin{equation}
	\begin{split}
		Y_{cr} = \text{APLR}_C(Y_c, M_c),
	\end{split}
\end{equation}
\begin{equation}
	\begin{split}
		Y_{ir} = \text{APLR}_I(Y_i, M_i),
	\end{split}
\end{equation}
where \( \text{APLR}_C(\cdot) \) and \( \text{APLR}_I(\cdot) \) denote the adaptive packet loss recovery functions for common and individual features, respectively. Additionally, CSI is incorporated to adapt packet loss recovery to varying signal-to-noise ratios, with detailed designs elaborated in subsequent subsections.

Subsequently, \( Y_{cr} \) and \( Y_{ir} \) are input into the Semantic Channel Decoder (SED) to restore semantic information precision:
\begin{equation}
	\begin{split}
		(Y_{cr}', Y_{ir}') = \text{SED}(Y_{cr}, Y_{ir}).
	\end{split}
\end{equation}

Next, \( Y_{ir}' \) is processed by the Individual Upsample Module to reconstruct the individual feature \( Y_{iru} \) with dimensions matching the common feature. The compressed spatio-temporal semantic information vector \( Y_{iru} + Y_{cr} \) is then input into the Spatial-Temporal Semantic Decoder to decompress and generate the spatial semantic information vector \( Y_s \) corresponding to the GOP:
\begin{equation}
	\begin{split}
		Y_{iru} = \text{UP}_I(Y_{ir}'),
	\end{split}
\end{equation}
\begin{equation}
	\begin{split}
		Y_s = \text{SD}_{ST}(Y_{iru} + Y_{cr}'),
	\end{split}
\end{equation}
\begin{equation}
	\begin{split}
		Y_s = \{Y_{s_1}, Y_{s_2}, \dots, Y_{s_n}\},
	\end{split}
\end{equation}
where \( \text{UP}_I(\cdot) \) denotes the individual feature upsampling function, and \( \text{SD}_{ST}(\cdot) \) represents the spatio-temporal semantic decoding function.

Finally, \( Y_s \) is separated into \( Y_{s_j} (j=1,2,\dots,n) \), and, combined with CSI, each is input into the Spatial Semantic Decoder to generate the corresponding reconstructed video frame \( Y_{g_j} \), yielding the reconstructed GOP \( Y_g \):
\begin{equation}
	\begin{split}
Y_{g_j} = \text{SD}_S(Y_{s_j}),
	\end{split}
\end{equation}
\begin{equation}
	\begin{split}
Y_g = \{Y_{g_1}, Y_{g_2}, \dots, Y_{g_n}\},
	\end{split}
\end{equation}
where \( \text{SD}_S(\cdot) \) denotes the function for reconstructing video frames from spatial semantic information vectors.

In the subsequent subsections, the proposed Semantic Mamba Block, Semantic Channel Codec, and Adaptive Packet Loss Recovery Module will be detailed. Other modules, similar to those in our previously published MSTVSC \cite{ref27}, utilize the 3D Swin Transformer as the backbone for semantic feature extraction and are not elaborated further here.

\subsection{The spatial semantic codec}
\begin{figure}[!htbp]
	\centering
	\includegraphics[width=0.5\textwidth]{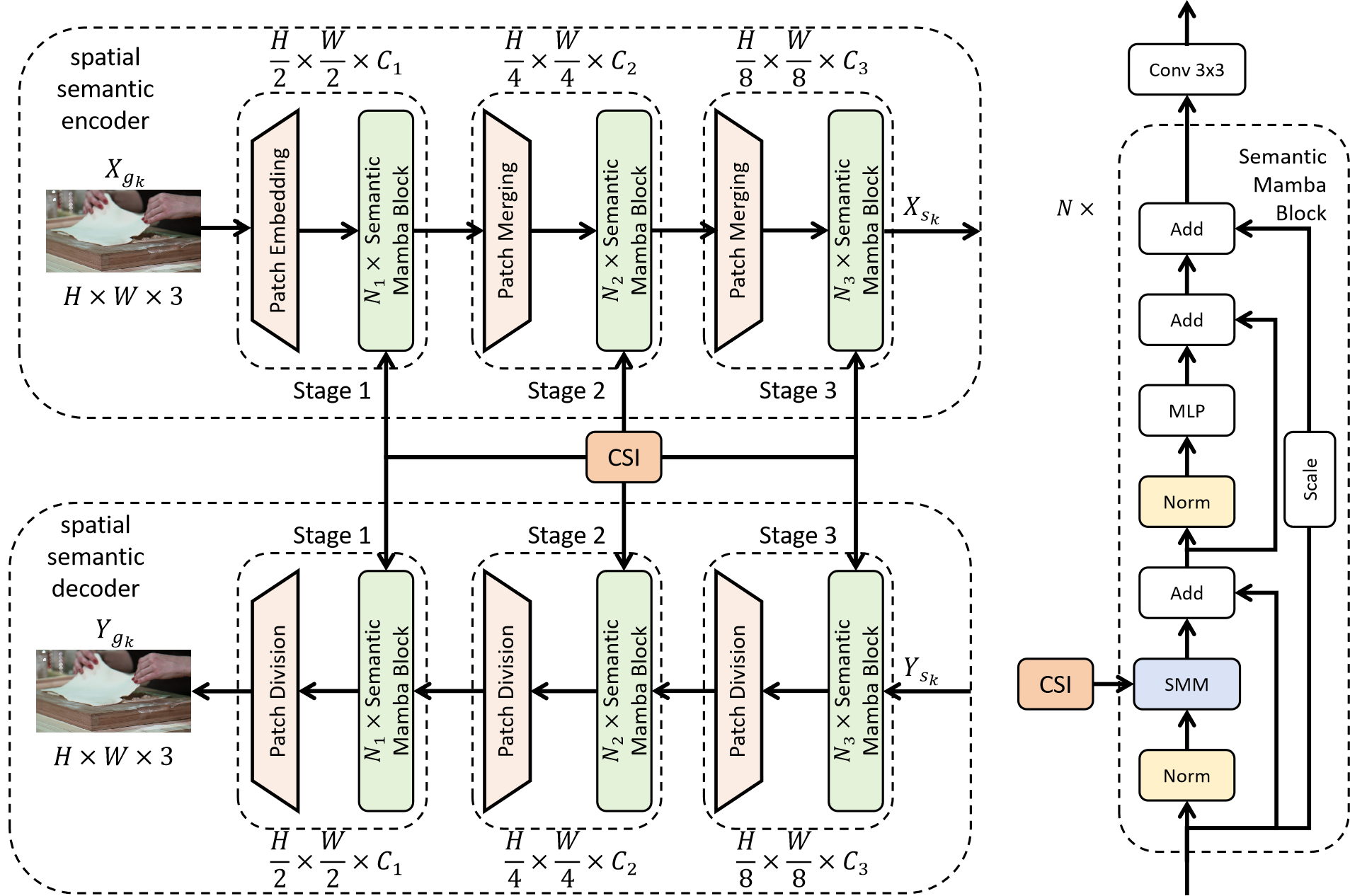}\\
	\caption{\footnotesize The architecture of the spatial semantic code.}
	\label{spatial semantic encoder}
\end{figure}
The network architecture of the Spatial Semantic Encoder-Decoder is depicted in Fig. \ref{spatial semantic encoder}, with its backbone consisting of Semantic Mamba Blocks regulated by CSI. In the spatial semantic encoder, a single frame \( X_{g_k} \in \mathbb{R}^{H \times W \times 3} \) from the GOP is first partitioned into \( H/2 \times W/2 \) non-overlapping patches. These patches are then processed by the Patch Embedding Module to generate initial patch embedding tokens. Subsequently, these tokens are fed into \( N_1 \) Semantic Mamba Blocks for further semantic information extraction. The network segment comprising the Patch Embedding Module and \( N_1 \) Semantic Mamba Blocks is collectively referred to as "Stage 1." The right side of Fig. \ref{spatial semantic encoder} illustrates the semantic extraction process of the Semantic Mamba Block, which integrates a Semantic Mamba Module with a Feed-Forward Network to process the image patches.

To enhance compression efficiency and extract more abstract and refined semantic information, patch merging layers and deeper network structures are incorporated. Specifically, the neighboring embedding outputs from Stage 1 are merged through the patch merging layer in Stage 2, producing concatenated embedding vectors of size \( 4C_1 \), which are subsequently reduced to size \( C_2 \) through dimensionality reduction. The resulting embedding tokens, with a resolution of \( H/4 \times W/4 \), are then input into \( N_2 \) Semantic Mamba Blocks for further processing. Stage 3 follows a similar structure, consisting of downsampling patch merging layers and \( N_3 \) Semantic Mamba Blocks. Through this progressively deepening network architecture, the model significantly enhances its capacity to capture long-range dependencies, leverage global information, and efficiently learn complex details in high-resolution images, ultimately generating the spatial semantic information vector \( X_{s_k} \) for a single frame.

In the spatial semantic decoder, the input is the spatial semantic information vector \( Y_{s_k} = Y_{iru} + Y_{cr} \) for a single frame. Each stage of the decoder is the inverse of the corresponding encoder stage, with its implementation mirroring that of the encoder, and thus is not elaborated further here.

\subsubsection{Semantic Mamba Block}
\begin{figure}[!htbp]
	\centering
	\includegraphics[width=0.45\textwidth]{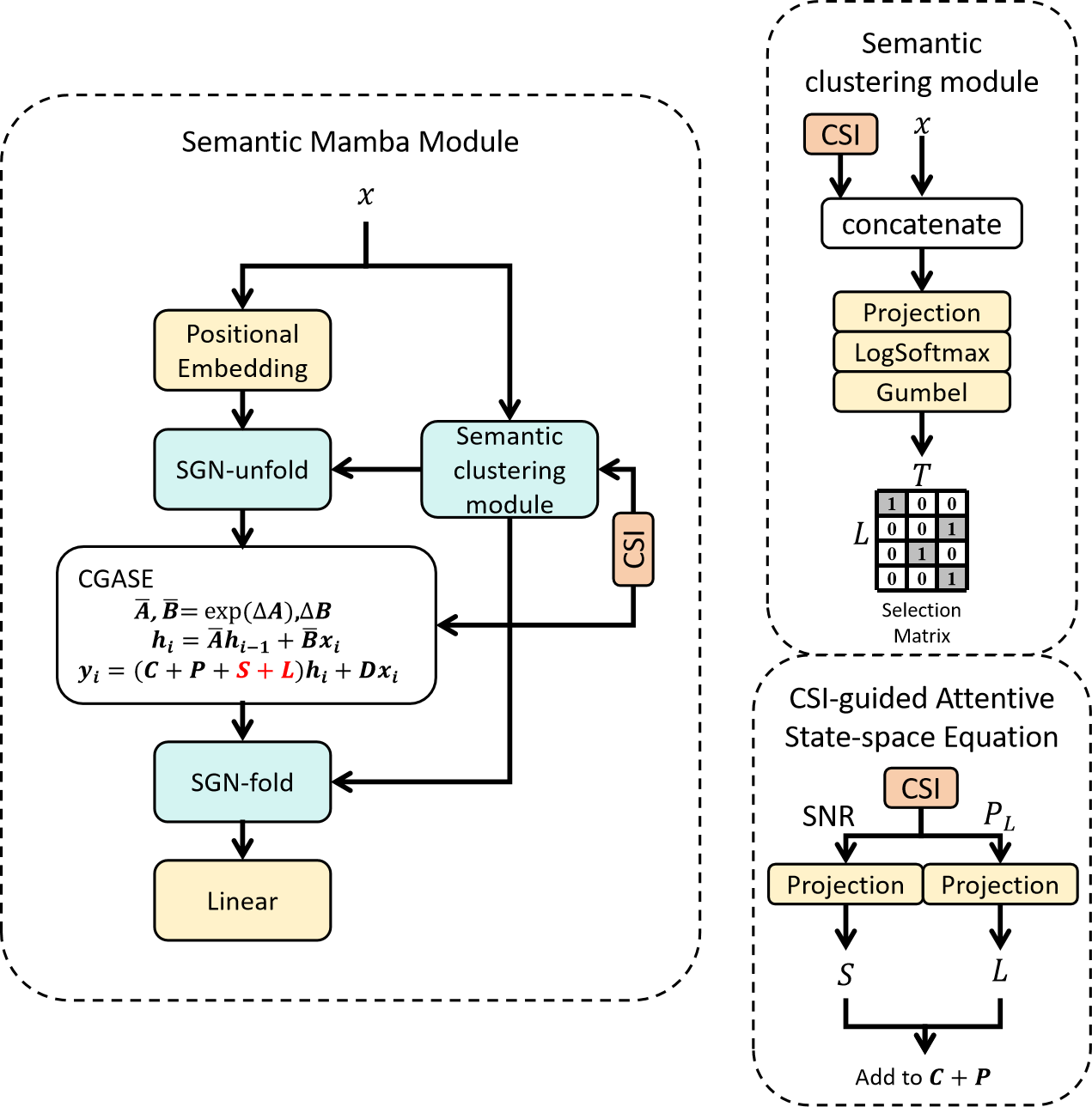}\\
	\caption{\footnotesize The architecture of the Semantic Mamba Module (SMM).}
	\label{ASE}
\end{figure}

The original state-space equations of Mamba are derived from state-space models (SSMs), conceptualized based on continuous systems, mapping a one-dimensional function or sequence \( x(t) \in \mathbb{R}^L \) to \( y(t) \in \mathbb{R}^L \) through a hidden state \( h(t) \in \mathbb{R}^N \). Formally, the dynamics of SSMs are described by the following ordinary differential equations (ODEs):
\begin{equation}
	\begin{split}
		h'(t) = \mathbf{A}h(t) + \mathbf{B}x(t),
	\end{split}
\end{equation}
\begin{equation}
	\begin{split}
		y(t) = \mathbf{C}h(t),
	\end{split}
\end{equation}
where \(\mathbf{A} \in \mathbb{R}^{N \times N}\) represents the system’s evolution matrix, and \(\mathbf{B} \in \mathbb{R}^{N \times 1}\), \(\mathbf{C} \in \mathbb{R}^{N \times 1}\) are projection matrices. In modern SSMs, this continuous ODE is approximated through discretization. Mamba, as a discretized version of the continuous system, introduces a time-scale parameter \(\Delta\), converting the continuous parameters \(\mathbf{A}\), \(\mathbf{B}\) into their discrete counterparts \(\overline{\mathbf{A}}\), \(\overline{\mathbf{B}}\). The transformation typically employs the zero-order hold (ZOH) method, defined as:
\begin{equation}
	\begin{split}
		\overline{\mathbf{A}} = \exp(\mathbf{\Delta A}),
	\end{split}
\end{equation}
\begin{equation}
	\begin{split}
		\overline{\mathbf{B}} = (\mathbf{\Delta A})^{-1} (\exp(\mathbf{\Delta A}) - \mathbf{I}) \cdot \mathbf{\Delta B},
	\end{split}
\end{equation}
\begin{equation}
	\begin{split}
		h_t = \overline{\mathbf{A}} h_{t-1} + \overline{\mathbf{B}} x_t,
	\end{split}
\end{equation}
\begin{equation}
	\begin{split}
		y_t = \mathbf{C}h_t.
	\end{split}
\end{equation}
Unlike traditional models that primarily rely on linear time-invariant SSMs, Mamba distinguishes itself by implementing a selective scanning mechanism (S6) as its core SSM operator. Building on this, it was pointed out in \cite{ref31} that Mamba’s inherent causal modeling limitation—where each token depends only on the previous token in the scanning sequence—restricts the full utilization of image pixels, posing new challenges for image reconstruction. To overcome the causality imposed by sequential scanning, a method was proposed in \cite{ref31} to incorporate prompts \(P\) into \(\mathbf{C}\), supplementing missing information from unscanned pixels. These prompts learn representations of pixel groups with similar semantics, achieving an attentive state-space recovery model with linear complexity, expressed as the Attentive State-Space Equation (ASE):
\begin{equation}
	\begin{split}
		h_i = \mathbf{A} h_{i-1} + \mathbf{B} x_i,
	\end{split}
\end{equation}
\begin{equation}
	\begin{split}
		\quad y_i = (\mathbf{C} + P) h_i + \mathbf{D} x_i.
	\end{split}
\end{equation}
Specifically, inspired by the ASE proposed in \cite{ref31}, the Semantic Mamba Module proposed in this paper is illustrated in Fig. \ref{ASE}. Initially, positional encoding is applied to the input \( x \) to preserve the original structural information. Subsequently, guided by CSI, the semantic clustering module processes \( x \), dividing it into several semantically similar groups. The Semantic-Guided Neighborhood (SGN) method is then employed to unfold the 2D image into a one-dimensional sequence, facilitating subsequent modeling with the CSI-Guided Attentive State-Space Equation. Finally, another SGN is used as the inverse operation of the previous SGN, refolding the sequence back into an image, followed by linear projection to obtain the module’s output.
\subsubsection{CSI-Guided Attentive State-Space Equation}

The objective is to modify the output matrix \( C \in \mathbb{R}^{L \times d} \), where \( L = HW \) represents the flattened image sequence length and \( d \) denotes the number of hidden states in Mamba, to achieve adaptive semantic information extraction based on channel conditions. To this end, a CSI-Guided Attentive State-Space Equation (CGASE) is proposed, building upon the ASE from Mambairv2 [31] but constrained by channel conditions. As shown in Fig. \ref{ASE}, the proposed CGASE dynamically adjusts the granularity of semantic extraction for a group of semantically similar pixels by incorporating CSI into \( C \). Specifically, the metrics SNR and packet loss rate \( P_L \), which quantify the severity of the two types of semantic distortions introduced by the wireless channel, are mapped to a \( d \)-dimensional vector through a linear layer. This vector is then added to the \( C + P \) term from the original ASE to control the influence of the hidden state \( h_i \) on the output, namely:
\begin{equation}
	\begin{split}
		\quad y_i = (\mathbf{C} + P+S+L) h_i + \mathbf{D} x_i.
	\end{split}
\end{equation}
By explicitly modeling channel distortions, CGASE enhances the model's robustness to semantic deviation and erasure, prioritizing the retention of critical semantic information that significantly impacts reconstruction metrics, thereby mitigating the effects of semantic distortions.

\subsubsection{Semantic clustering module}
The causal modeling characteristic of Mamba leads to the adverse effect of long-distance attenuation. To address this, a SGN approach was proposed in \cite{ref31}, enabling semantically similar tokens to be spatially closer in the unfolded sequence. However, in wireless communication, the transmission of semantic information is subject to interference from channel-induced semantic distortions. Without incorporating CSI, the model would attempt to account for all possible semantic distortion scenarios during training, resulting in semantic clustering labels for each patch that do not adapt to improving or deteriorating channel conditions. This lack of flexibility leads to suboptimal granularity: labels remain neither finer under improved channel conditions nor coarser under severe conditions. Particularly in the semantic decoder, strong semantic distortions may cause patches that are originally semantically similar to exhibit significant differences, potentially leading to misclassification into incorrect semantic clustering labels. Therefore, CSI is introduced to adjust the tolerance of patch classification, thereby enhancing robustness.

Based on this concept, the Semantic Clustering Module is employed to control the SGN-unfold process. This module groups pixels belonging to the \( i \)-th prompt category into the \( i \)-th semantic group and then combines different groups based on the category value \( i \), generating a semantic neighborhood sequence. Subsequently, this sequence is input into the proposed CGASE for state-space modeling. Finally, SGN-fold, the inverse transformation of SGN-unfold, is applied to reshape the semantic space sequence back into a spatial feature map, yielding the output.

\subsection{Semantic Channel Codec Based on Dynamic Semantic clustering centers}
\begin{figure}[!htbp]
	\centering
	\includegraphics[width=0.5\textwidth]{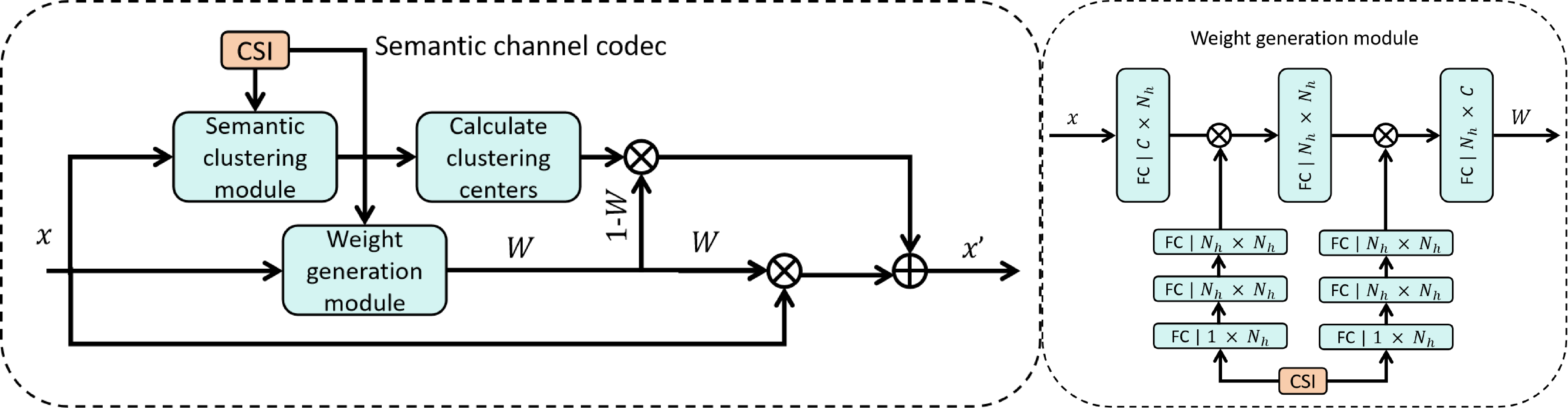}\\
	\caption{\footnotesize The architecture of the Semantic Channel Codec.}
	\label{spatial time semantic encoder}
\end{figure}

As noted in \cite{ref28}, the trade-off learning between characteristic and common information is a key characteristic of semantic communication, aligning with the synonymous mapping concept proposed in \cite{ref29}. Consequently, the robustness of semantic information to distortions can be adjusted by explicitly controlling the distance between semantic vectors (characteristic features) and Semantic clustering centers (common information).
\begin{figure}[!htbp]
	\centering
	\includegraphics[width=0.5\textwidth]{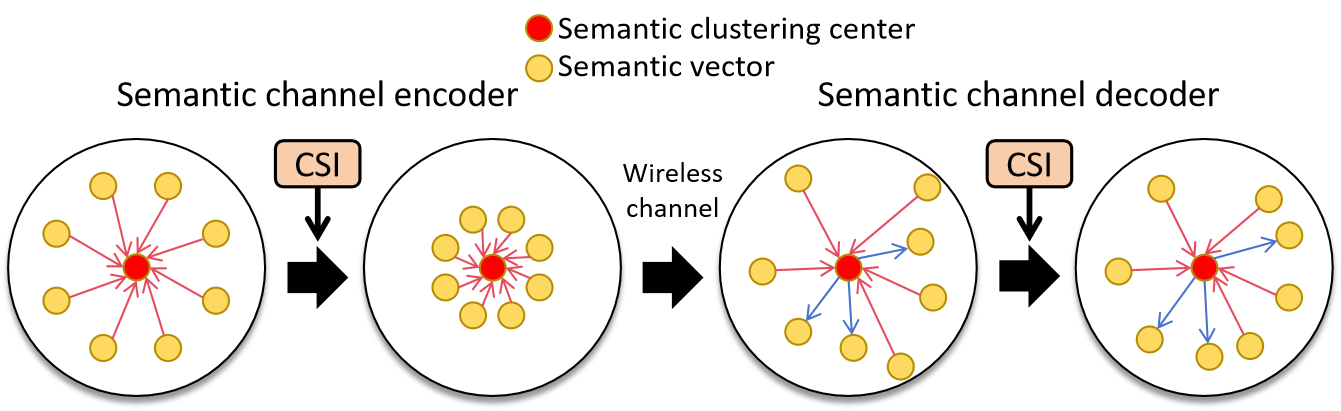}\\
	\caption{\footnotesize Diagram of the Semantic Channel Codec Functionality. In the figure, the red arrow represents convergence toward the semantic clustering center, while the blue arrow represents divergence from the semantic clustering center.}
	\label{star}
\end{figure}
The proposed semantic channel codec is illustrated in Fig. \ref{spatial time semantic encoder}, taking semantic information \( x \) as input. Similar to the Semantic Mamba Block, the semantic channel codec employs a Semantic Clustering Module. Building on this, the codec calculates the clustering centers for each semantic cluster, which serve as the Semantic clustering centers \( S_b \) for the respective clusters. Since the video frames within each GOP vary, the computed \( S_b \) constitutes dynamic Semantic clustering centers that change with each GOP. Additionally, the semantic information \( x \) is fused with the input CSI and processed through a weight generation module to produce a weight \( W \), which controls the distance between \( x \) and \( S_b \). The output of the semantic channel codec, \( x' \), is expressed as:
\begin{equation}
	\begin{split}
		x' = Wx + (1-W)S_b.
	\end{split}
\end{equation}

Fig. \ref{star} illustrates the effect of the semantic channel codec in a wireless channel. At the semantic channel encoder, semantic vectors within the same cluster are used to compute the corresponding Semantic clustering center, and the distance between the semantic vectors and the Semantic clustering center is adjusted based on CSI to enhance robustness. Lower signal-to-noise ratios result in semantic vectors being closer to the Semantic clustering center. After transmission through the wireless channel, the semantic vectors are subject to varying degrees of distortion. The semantic channel decoder then adjusts the semantic vectors based on CSI, striving to restore the original semantic information.

\subsection{Adaptive Packet Loss Recovery Module}
\begin{figure}[!htbp]
	\centering
	\includegraphics[width=0.49\textwidth]{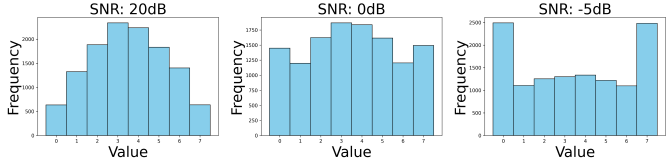}\\
	\caption{\footnotesize Histogram of Received Semantic Information Distribution under Different SNRs.}
	\label{fenbu}
\end{figure}

Due to semantic deviation introduced by the wireless channel, a packet loss recovery module without adaptive capabilities cannot accommodate a wide range of SNR variations. This is attributed to the significant changes in the distribution of received semantic information under different SNR conditions. Fig. \ref{fenbu} illustrates the distribution of received semantic information using the sDMCM \cite{Teng} modulation method at the receiver under varying SNR conditions. It can be observed that at high SNR, the semantic information distribution approximates a Gaussian distribution, primarily concentrated around values 3 and 4. As SNR decreases, the distribution shifts toward the extremes, with the proportions of values 0 and 7 increasing rapidly. This is because lower signal-to-noise ratios (SNRs) cause more severe deviations of the received constellation points from their transmitted positions after mapping the original information onto the constellation. At an SNR of 20 dB, the received constellation points can be considered essentially indistinguishable from the transmitted information distribution. However, at an SNR of -5 dB, the majority of received constellation points exhibit severe deviations, exceeding the maximum and minimum values of the original constellation diagram, and are thus clipped to 0 or 7. Consequently, the packet loss recovery module must adapt to different received semantic information distributions based on SNR.
\begin{figure}[!htbp]
	\centering
	\includegraphics[width=0.4\textwidth]{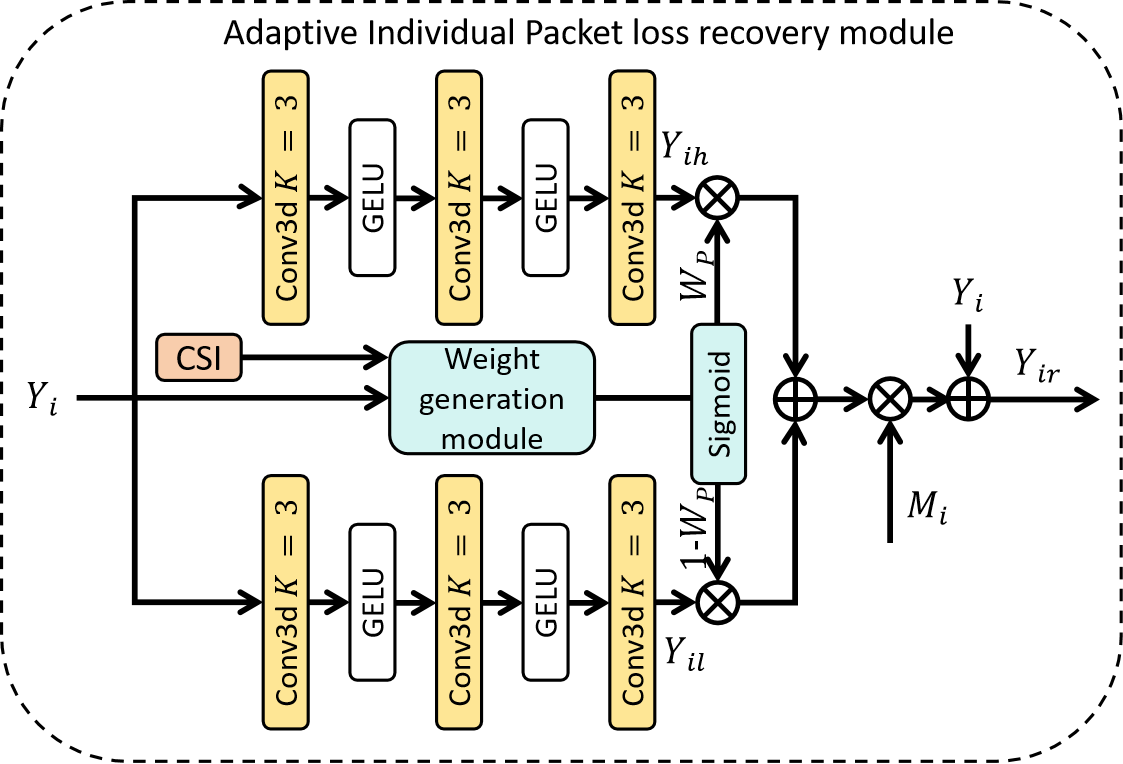}\\
	\caption{\footnotesize The architecture of the Adaptive Packet Loss Recovery Module.}
	\label{APM}
\end{figure}

The proposed Adaptive Individual Packet Loss Recovery Module is depicted in Fig. \ref{APM}, where the convolutional-based information recovery module is divided into two paths. The upper path processes \( Y_i \) through multiple convolutional layers to produce \( Y_{ih} \), which is trained under high SNR conditions. Similarly, the lower path produces \( Y_{il} \), trained under low SNR conditions. The middle path processes \( Y_i \) through a CSI-controlled weight generation module, followed by a sigmoid function to constrain the output to [0, 1], generating a weight \( W_P \) that is used to weight \( Y_{ih} \) and \( Y_{il} \). The weighted result is then multiplied by the semantic erasure matrix \( M_i \) to retain only the values at positions where semantic information is erased, which are subsequently added to the received semantic information to complete the adaptive packet loss recovery. This process can be expressed as:
\begin{equation}
	\begin{split}
		Y_{ir} = (W_P Y_{ih} + (1-W_P) Y_{il}) M_i + Y_i.
	\end{split}
\end{equation}

\section{Experimental Results}
\subsection{Experimental Setup}
\subsubsection{Datasets}
The proposed MamVSC model was trained on the Vimeo-90k dataset \cite{ref32}, which contains 89,800 video clips, each consisting of 7-frame sequences. The Group of Pictures (GOP) size was set to 4, and each frame within a GOP was randomly cropped to 256 × 256 pixels. Subsequently, the frame data was normalized to the range [0, 1]. After training, the model's performance was evaluated using the HEVC test dataset \cite{ref33} and the UVG dataset \cite{ref34}. To better demonstrate performance at high resolutions, the following subsets were selected: Class B (1920 × 1080), and UVG (3840 × 2160). These video subsets not only feature higher resolutions but are also more common in practical applications, imposing greater demands on communication systems.

\subsubsection{Comparison Schemes}
The proposed MamVSC scheme is compared with the following approaches: the Swin Transformer-based MSTVSC scheme \cite{ref27}, the CNN-based MDVSC scheme \cite{ref22}, and classical separate source-channel coding schemes, including H.264 \cite{ref35} and H.265 \cite{ref36} for source coding, combined with LDPC \cite{ref37} for channel coding. All semantic communication schemes employ sDMCM \cite{Teng} modulation for transmission. In this study, a 1/2 LDPC code indicates that half of the total code length is used for redundancy. Additionally, ablation studies were conducted. The experiments in this work are conducted under a no-retransmission policy, where every packet is sent exactly once.
\subsubsection{Evaluation Metrics}
The end-to-end image transmission performance of the proposed MamVSC model and the comparison schemes was evaluated using widely adopted pixel-wise metrics, PSNR, and perceptual metric, MS-SSIM \cite{ref38}. Additionally, the Channel Bandwidth Ratio (CBR) \cite{ref14} is a widely used metric for measuring wireless channel bandwidth overhead in the field of semantic communication, as well as in semantic digital communication, as evidenced by \cite{ref16, ref22, ref27, Qi2024, Dai2023}. Therefore, in this paper, CBR is used to measure bandwidth utilization efficiency, calculated as:

\begin{equation}
	\begin{split}
		\text{CBR} = \frac{v}{l},
	\end{split}
\end{equation}
where $ l $ is the dimension of the source data $ X_g $, and $ v $ is the sum of the dimensions of the vectors $ X_c $ and $ X_i $ after mapping by the semantic encoder, also referred to as the channel bandwidth cost. Notably, since this study focuses on digital communication systems, semantic data transmission requires quantization into bits. All semantic models adopt 3-bit quantization.

\subsection{Results Analysis}

\subsubsection{The Relationship between Semantic Erasure Probability and SNR}

\subsubsection{Performance at Different CBRs}
\begin{figure}[!htbp]
	\centering
	\includegraphics[width=0.5\textwidth]{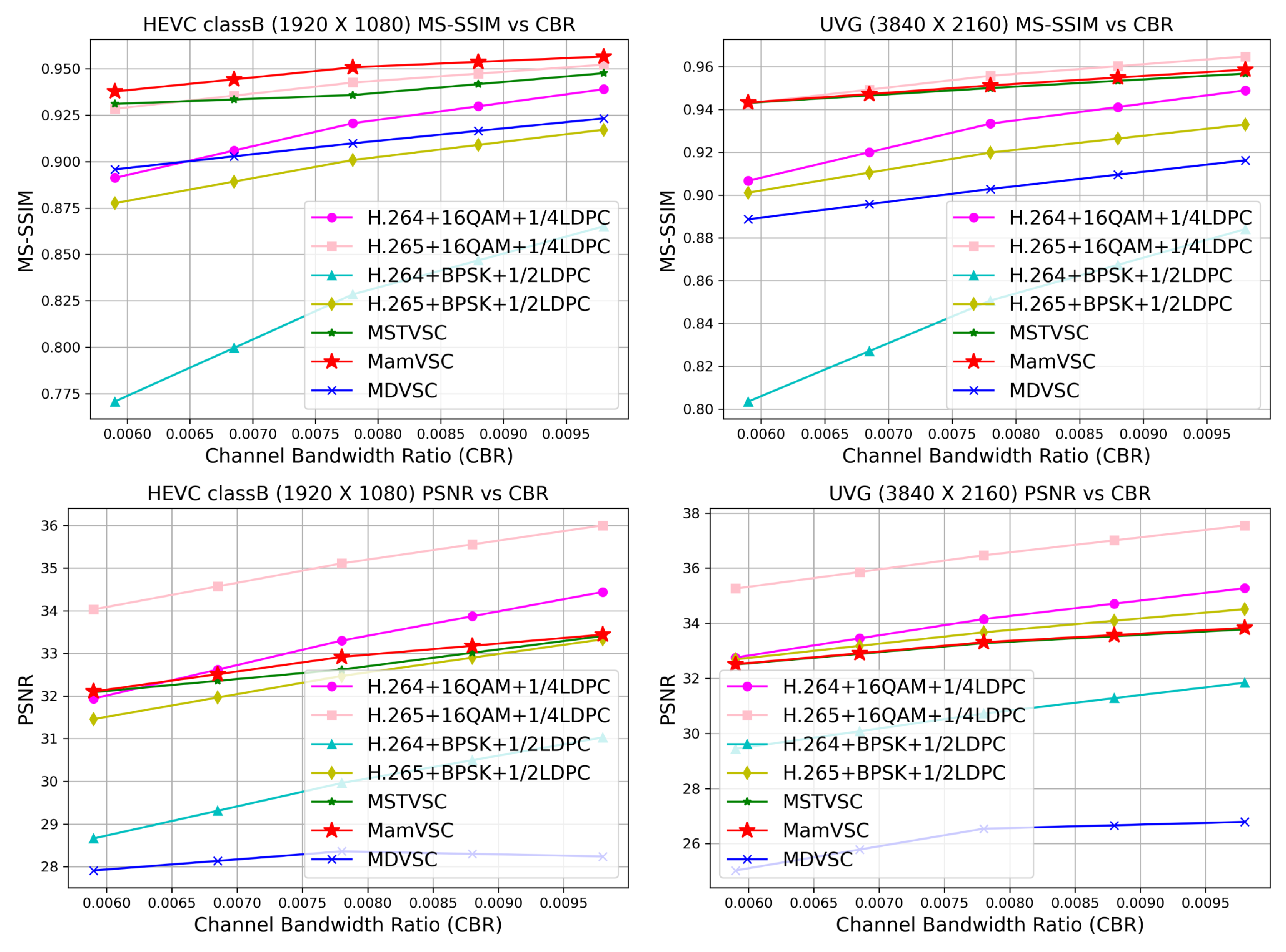}\\
	\caption{\footnotesize The relationship between PSNR and MS-SSIM performance of different encoding-decoding schemes and CBR under error-free transmission is depicted.}
	\label{CBR}
\end{figure}

Fig. \ref{CBR} presents the video reconstruction metrics under error-free transmission for different CBRs. In terms of PSNR, the proposed MamVSC outperforms the channel-robust combination of H.264 with BPSK + 1/2 LDPC and slightly surpasses H.265 with BPSK + 1/2 LDPC, except at the 3840 × 2160 resolution. Compared to other video semantic communication systems, MamVSC significantly outperforms MDVSC and slightly outperforms MSTVSC. Regarding MS-SSIM, MamVSC demonstrates clear superiority over the channel-robust combinations of H.264 and H.265 with BPSK + 1/2 LDPC, surpasses the transmission-efficient combination of H.264 with 16QAM + 1/4 LDPC, and outperforms H.264 with the transmission-efficient combination at 1920 × 1080 resolution. Additionally, it narrows the gap with H.265 with 16QAM + 1/4 LDPC at 3840 × 2160 resolution. This is attributed to the MS-SSIM metric, which evaluates perceptual quality more aligned with human perception, making it more suitable for assessing semantic communication performance.

Compared to other video semantic communication systems, MamVSC exhibits superior performance over MDVSC in both PSNR and MS-SSIM metrics and slightly outperforms MSTVSC.

\begin{figure*}[!htbp]
	\centering
	\includegraphics[width=0.9\textwidth]{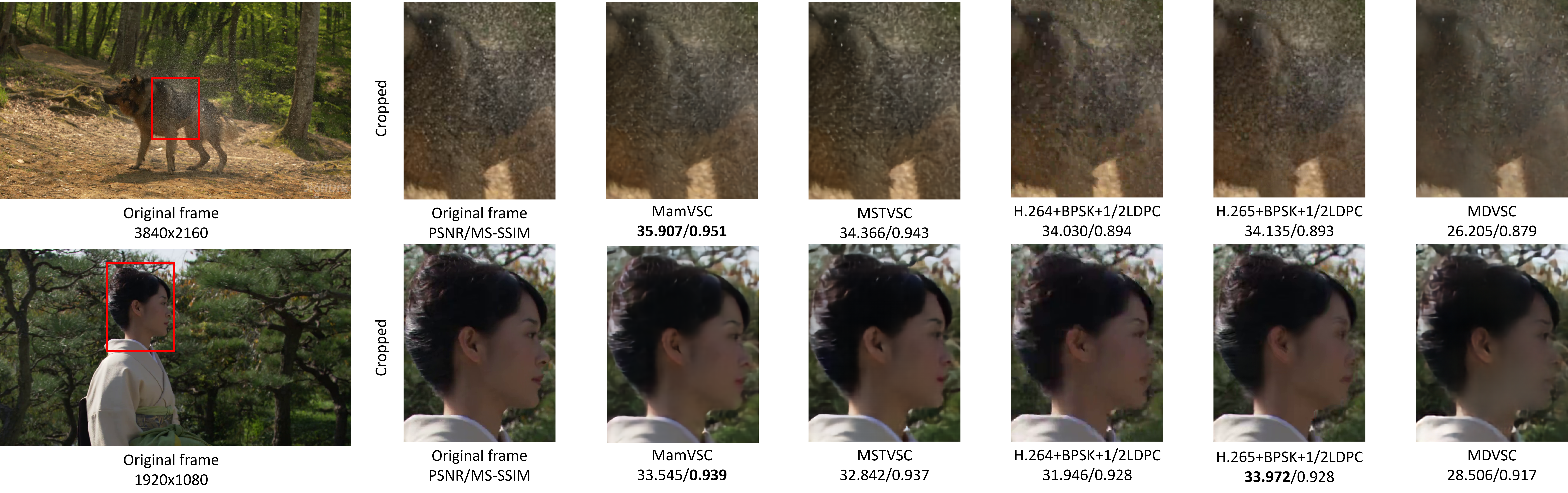}\\
	\caption{\footnotesize An example of visual comparison is provided, with the CBR set to 0.0078. For the first two rows, the first column displays the original frame, the second column shows a cropped patch from the original frame, and the third to sixth columns present the reconstructed frames using different schemes as shown in Fig. \ref{CBR}, with subtitles indicating PSNR/MS-SSIM values. The first row demonstrates performance at Full HD resolution, while the second row shows performance at 4K Ultra HD resolution.}
	\label{visualization_CBR}
\end{figure*}

Fig. \ref{visualization_CBR} illustrates the visualized reconstruction results of different schemes under various resolutions, with the CBR uniformly set to 0.0078 and the channel coding and modulation scheme for H.264/H.265 set to 1/2 LDPC and BPSK. In the first row, at a resolution of 3840 × 2160, as shown in the cropped and magnified images on the right, water droplets splashed by a dog's movement exhibit perceptible pixelation and slight color noise in H.264/H.265. In MDVSC, the water droplets and the dog's fur blend together, resulting in poor sharpness. In contrast, the reconstructed frames of MamVSC and MSTVSC clearly distinguish the water droplets and the dog's fur, with MamVSC achieving more accurate color reproduction, leading to the highest PSNR and MS-SSIM scores. In the second row, at a resolution of 1920 × 1080, the cropped and magnified images reveal that traditional coding methods exhibit pixelation and artifacts, with dynamic blurring on human faces. In MDVSC, the image color shows noticeable deviation, appearing overly gray. MamVSC and MSTVSC avoid artifacts and dynamic blurring in this comparison, with MamVSC demonstrating more accurate color reproduction, resulting in the highest MS-SSIM score.

\subsubsection{Performance at Different Semantic Distortion}

\begin{figure}[!htbp]
	\centering
	\includegraphics[width=0.49\textwidth]{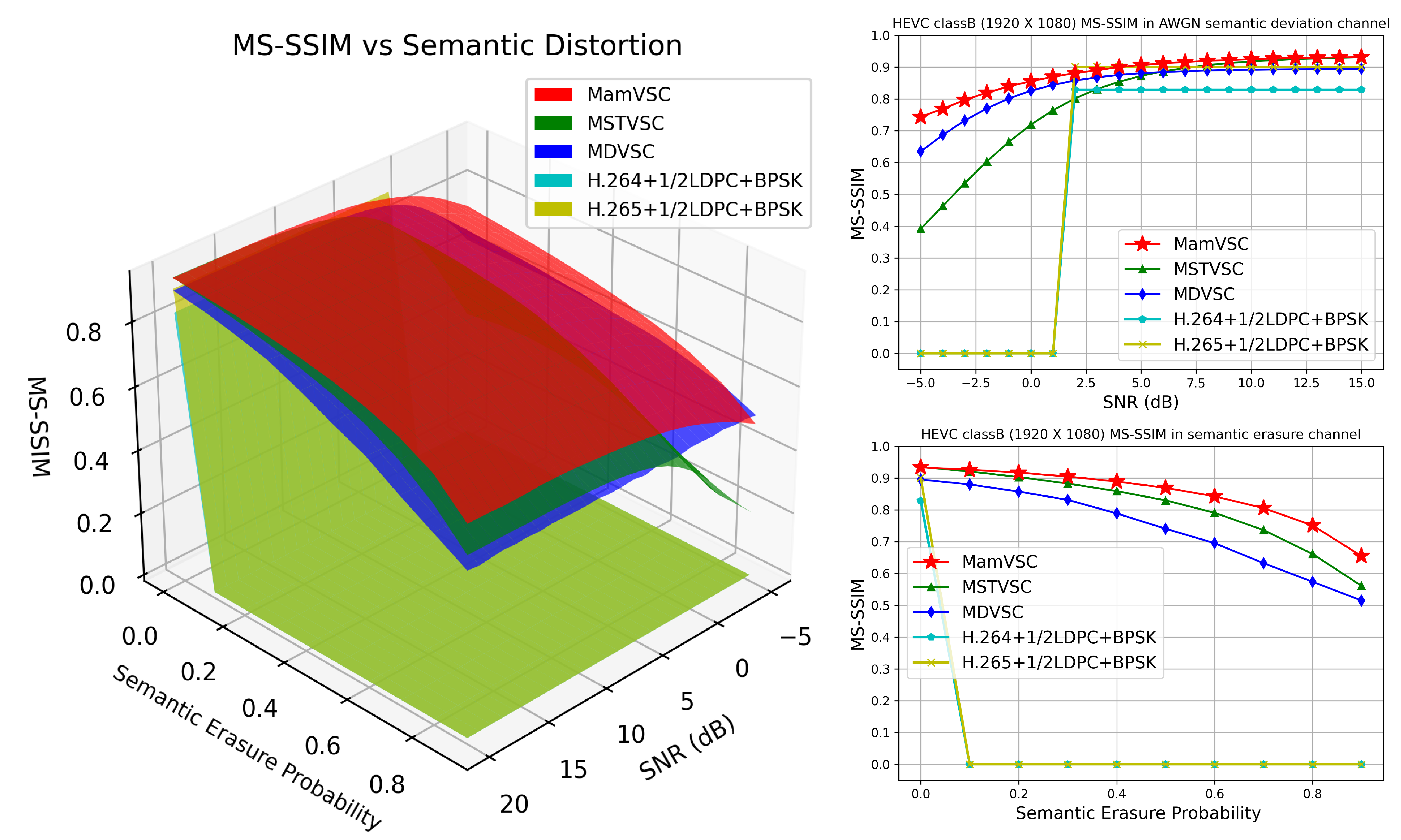}\\
	\caption{\footnotesize The three-dimensional surfaces plot illustrating the variation of semantic performance metrics MS-SSIM with SNR and semantic erasure probability in AWGN channel is presented. Additionally, performance curves of MS-SSIM versus SNR are shown for the case of zero semantic erasure probability, together with MS-SSIM versus semantic erasure probability under a fixed SNR of 20 dB.}
	\label{3d2}
\end{figure}
\begin{figure}[!htbp]
	\centering
	\includegraphics[width=0.49\textwidth]{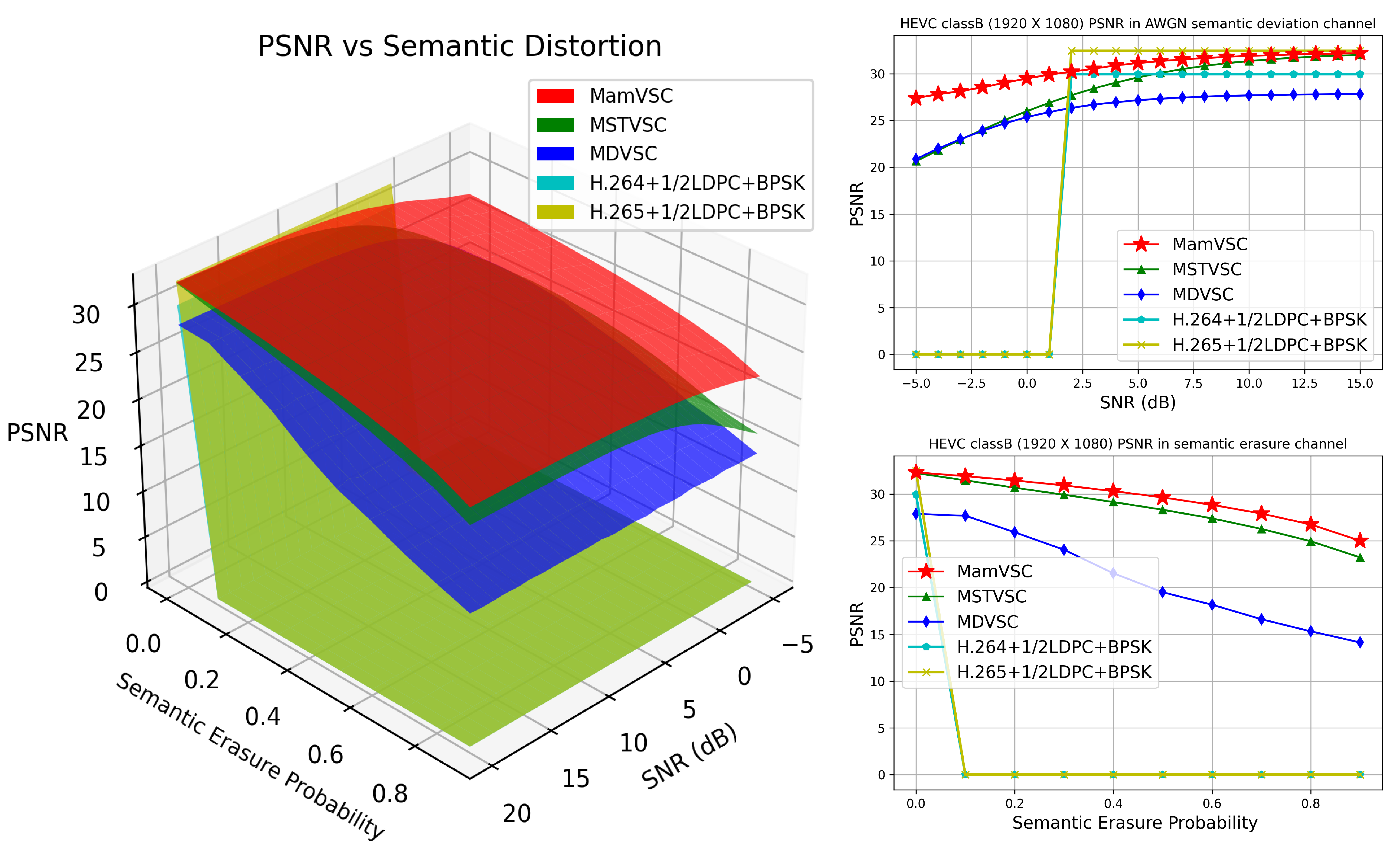}\\
	\caption{\footnotesize The three-dimensional surfaces plot illustrating the variation of semantic performance metrics PSNR with SNR and semantic erasure probability in AWGN channel is presented. Additionally, performance curves of PSNR versus SNR are shown for the case of zero semantic erasure probability, together with PSNR versus semantic erasure probability under a fixed SNR of 20 dB.}
	\label{3d3}
\end{figure}
\begin{figure}[!htbp]
	\centering
	\includegraphics[width=0.5\textwidth]{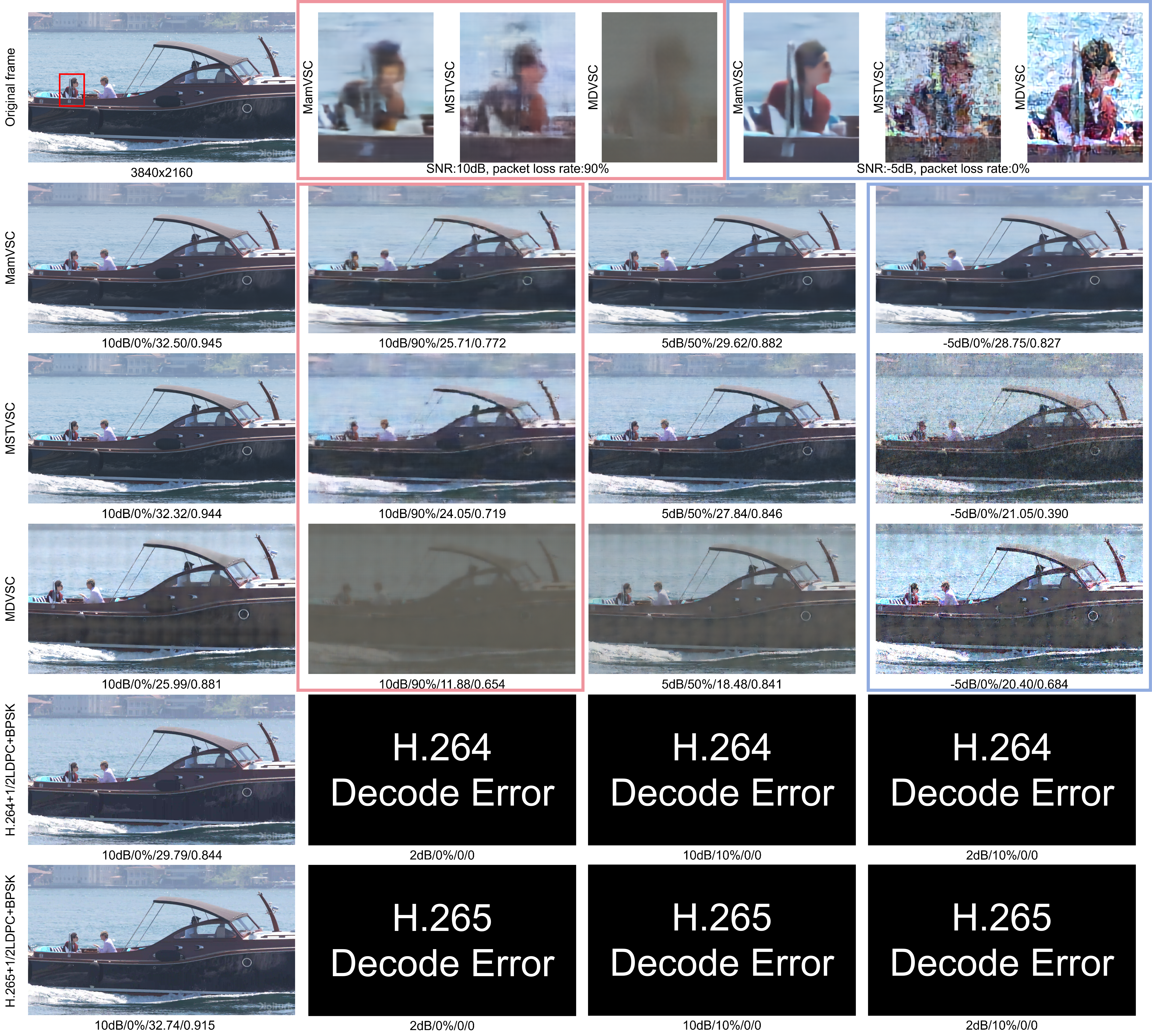}\\
	\caption{\footnotesize Visual comparison of reconstructed frames under different semantic distortions at CBR = 0.0078. The first row shows the original frame (leftmost column) and cropped/enlarged regions of reconstructed frames obtained by MamVSC, MSTVSC, and MDVSC under various semantic distortion conditions (columns 2–7). Colored bounding boxes indicate corresponding regions shown in the full-size frames below. Rows 2–5 display the complete reconstructed frames using: MamVSC, MSTVSC, MDVSC, H.264 + 1/2 LDPC + BPSK and H.265 + 1/2 LDPC + BPSK. The subtitles indicate SNR/the packet loss/PSNR/MS-SSIM values.}
	\label{packet_loss_v}
\end{figure}

Fig. \ref{3d2} and Fig. \ref{3d3} illustrates the robustness of different video semantic communication systems and traditional communication systems against semantic deviation (channel noise) and semantic erasure (packet loss). It can be observed that the proposed MamVSC significantly outperforms MSTVSC and MDVSC in terms of PSNR robustness for both types of semantic distortions, demonstrating its superior robustness in color accuracy.  The limited PSNR robustness of MDVSC aligns with the visualization results in Fig. \ref{packet_loss_v}, where severe color deviation is evident. Additionally, in terms of MS-SSIM performance, MamVSC surpasses MSTVSC and MDVSC across the entire test range. Fig. \ref{3d2} and Fig. \ref{3d3} indicates that designing a semantic communication system that addresses only one type of semantic distortion leads to insufficient robustness against the other. For instance, MDVSC's performance rapidly degrades as semantic erasure increases, while MSTVSC's performance declines quickly as semantic deviation intensifies. Furthermore, the conventional H.264/H.265 + 1/2 LDPC + BPSK scheme, due to its reliance on highly context-adaptive entropy coding and predictive differential structures, requires error-free bit transmission to function correctly. As a result, packet loss or SNR dropping below the LDPC error-correction threshold leads to complete decoding failure and zero performance—a well-known phenomenon referred to as the cliff effect.

Fig. \ref{packet_loss_v} illustrates the reconstructed image visualization results of different schemes under various semantic distortion conditions. Three representative extreme scenarios are particularly highlighted and enlarged: cases dominated by semantic erasure and cases dominated by semantic deviation pronounced.

Overall, the proposed MamVSC consistently exhibits superior reconstruction performance compared to all baseline schemes across the tested communication conditions.

In the extreme scenario dominated by semantic erasure (SNR = 10 dB, packet loss rate = 90\%, corresponding to the red-boxed region in Fig. 14), both MamVSC and MSTVSC (specially designed to combat semantic erasure distortion) effectively preserve image structure and semantic integrity. In contrast, MDVSC (optimized solely against semantic deviation) suffers severe performance degradation, manifested as significant overall distortion, pronounced graying, and substantial loss of fine details.

In the scenario dominated by semantic deviation (SNR = -5 dB, packet loss rate = 0\%, corresponding to the blue-boxed region in Fig. 14), MamVSC demonstrates clearly superior reconstruction quality over both MSTVSC and MDVSC. The reconstructed images exhibit almost no visible mosaicking, with only mild blurring, achieving objective quality metrics of MS-SSIM  $>$ 0.8 and PSNR  $>$ 28 dB. Benefiting from its dedicated design and training against semantic deviation, MDVSC slightly outperforms MSTVSC, yet still shows noticeable color shift and block artifacts. MSTVSC, having not accounted for semantic deviation, produces the most severe blurring.

Compared with the traditional video coding scheme (H.264/H.265 + 1/2 LDPC + BPSK), MamVSC exhibits significantly stronger robustness in low-SNR and high-packet-loss regimes, maintaining basic visual readability even under extreme channel conditions. The traditional scheme, however, displays a pronounced cliff effect: once the SNR falls below the LDPC decoding threshold or packet loss occurs (without retransmission), the video becomes completely undecodable, leading to total collapse of the reconstructed content.

\begin{figure}[!htbp]
	\centering
	\includegraphics[width=0.5\textwidth]{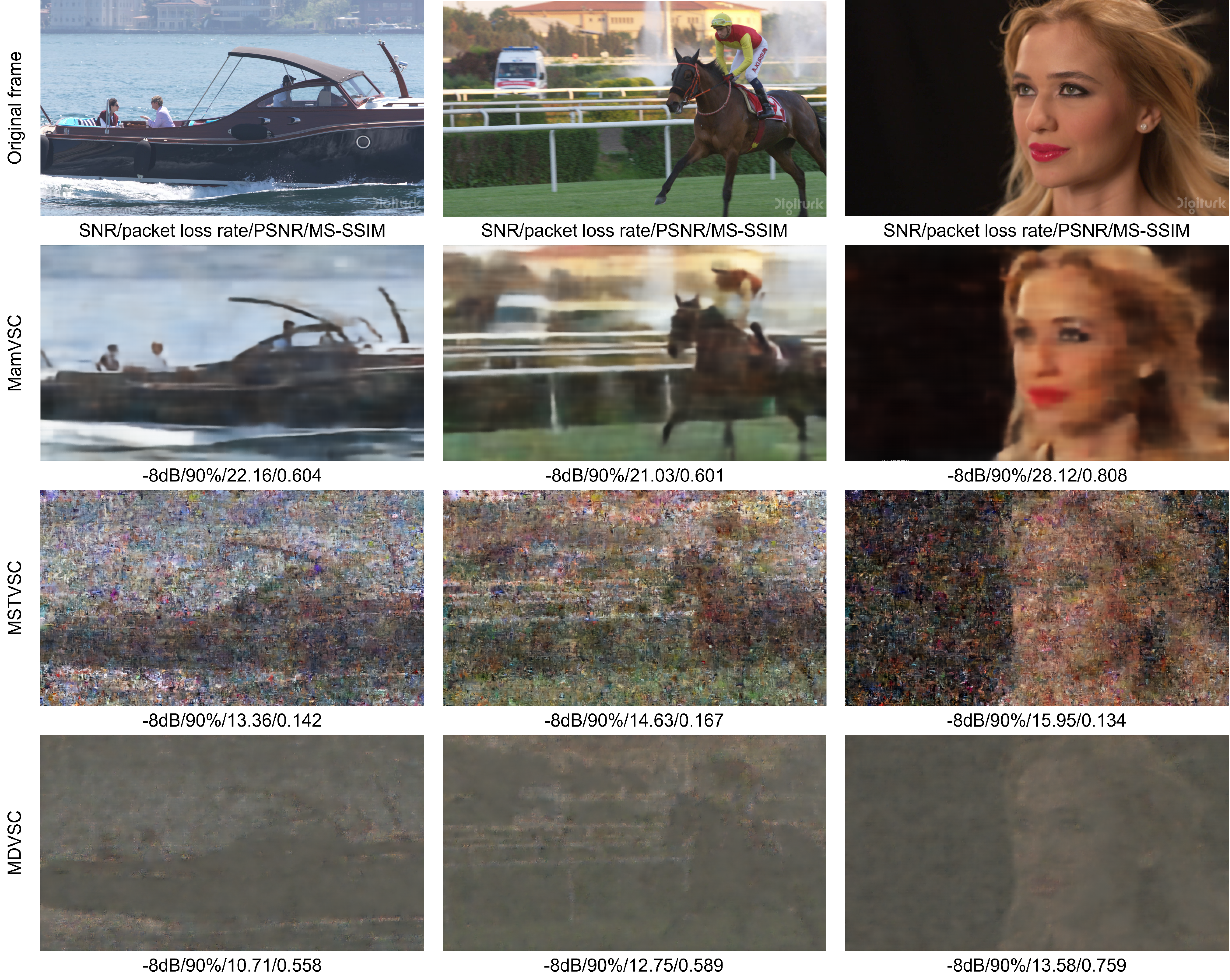}\\
	\caption{\footnotesize Visualization results under extreme semantic distortion conditions (SNR = -8 dB, packet loss rate = 90\%).	The first row shows the original frame. The second to fourth rows present the reconstructed video frames obtained by MamVSC, MSTVSC, and MDVSC, respectively. The subtitles indicate SNR/the packet loss/PSNR/MS-SSIM values.}
	\label{packet_loss_8}
\end{figure}

Fig. \ref{packet_loss_8} illustrates a scenario of extreme semantic distortion, where SNR = -8 dB and the packet loss rate reaches 90\%. Under such harsh conditions, the transmitted semantic information is severely corrupted and largely erased. Even in this extremely challenging setting, the MamVSC scheme is still able to preserve minimal recognizable semantic and structural information in the reconstructed video frames. The hull of the ship and the people on board remain identifiable, the dynamic postures of the horses and riders are basically recoverable, and the facial features of the woman can still be discerned. The overall reconstruction quality achieves PSNR $>$ 21 dB and MS-SSIM $>$ 0.6. In contrast, both MSTVSC and MDVSC suffer from severe distortion, exhibiting prominent colored noise patches and pervasive graying artifacts, resulting in almost complete loss of semantic content and rendering the reconstructed frames nearly unrecognizable.

\section{Conclusion}

This paper proposes MamVSC, a Mamba-based semantic wireless video communication system that effectively addresses both semantic deviation and semantic erasure through a novel semantic grouping mechanism. By incorporating channel state information as input at both the transmitter and receiver sides, MamVSC enables dynamic adaptation to varying channel conditions, thereby achieving robust and better performance across diverse semantic distortion scenarios.
To realize this goal, we design a CSI-Guided Attentive State-Space Equation module that adaptively regulates the granularity of semantic feature extraction in the codec according to CSI. Furthermore, a dynamic semantic channel encoding module is introduced, which controls transmission robustness by adjusting the distance between semantic vectors and their corresponding clustering centers. A dynamic adaptive packet loss recovery mechanism is also proposed to flexibly handle varying patterns of received semantic information, significantly enhancing system applicability and resilience.
Extensive experimental comparisons with conventional video codecs (H.264 and H.265) and semantic communication methods (MSTVSC and MDVSC) demonstrate that MamVSC consistently achieves superior video reconstruction quality while exhibiting markedly higher robustness against both semantic deviation and semantic erasure under challenging wireless conditions.
These results highlight the strong potential of Mamba-based architectures combined with CSI-aware semantic adaptation for next-generation wireless video communication systems.

%
%
%
%
%


\end{document}